\documentclass[twocolumn,showpacs,preprintnumbers,amsmath,amssymb]{revtex4}
\usepackage{amsfonts,amssymb}
\usepackage{times,mathptm}
\usepackage{amsmath,amsthm,graphicx}
\usepackage{dcolumn}
\usepackage{bm}
\usepackage{hyperref}

\begin{document}

\preprint{APS/123-QED}

\title{Linear Depth Stabilizer and Quantum Fourier Transformation Circuits with \\
no Auxiliary Qubits in Finite Neighbor Quantum Architectures}

\author{Dmitri Maslov}
\affiliation{
Institute for Quantum Computing, University of Waterloo, Waterloo, ON, Canada, N2L 3G1. Email: 
\href{mailto:dmitri.maslov@gmail.com}{dmitri.maslov@gmail.com}.\\
}

\date{\today}

\begin{abstract}
In this paper we investigate how quantum architectures affect the
efficiency of the execution of the Quantum Fourier Transform (QFT) and 
linear transformations, which are essential parts of the stabilizer/Clifford group circuits.
In particular, we show that in most common and realistic physical architectures 
including Linear Nearest Neighbor (LNN), 2D lattice, and bounded 
degree graph (containing a chain of length $n$), $n$-qubit QFT and 
$n$-qubit stabilizer circuits can be parallelized to 
linear depth using no auxiliary qubits. 
We construct lower bounds that show the efficiency 
of our approach. 
\end{abstract}


\maketitle

\section{Introduction}\label{sec:back}
Quantum computation has attracted attention because it appears to 
reduce the computational complexity of certain calculations, see, for example, \cite{co:g,ar:s}.
For the quantum circuit model of computation, there exists a number of 
physical quantum information processing implementations, such as liquid NMR 
(up to 12 qubits at a time) \cite{ar:nmrd}, and trapped ions (8 qubits) \cite{ar:s-k}. Generally, a 
large number of qubits is required for computational purposes. In this work,
we do not allow for any auxiliary qubits to be used in order to reflect the apparent hardness 
of scaling up quantum information processing devices. 

Quantum circuits have been optimized to require less space, fewer gates and 
smaller depth. This is important from the point of view of the efficient 
potential realization of the quantum algorithms. As discussed in the first 
paragraph, we address the issue of space minimization by restricting the 
number of auxiliary qubits to zero. Our next focus is on depth minimization.
This is because a small depth circuit does not only mean a fast computation, but also
helps reduce the effect of decoherence. For instance, it is possible 
to construct realistic examples in which a smaller depth circuit will require fewer levels 
of error correction, and each error correction code concatenation step is a very 
expensive operation \cite{bk:nc}. 

In this paper, the depth of a circuit is defined as the number of logic levels in it. 
Each logic level is a set of non-intersecting ``elementary'' gates. It is generally 
accepted that in a practical quantum information processing approach, it should be 
possible to execute independent gates in parallel. The gate libraries considered in 
the relevant literature include a set of single-qubit and CNOT gates (which is most 
likely an artifact of the well known result showing the completeness of this gate set, however,
CNOT may not necessarily be a natural gate for some quantum information processing proposals),
and any two-qubit operation. Indeed, given a Hamiltonian, any two-qubit operation can be 
efficiently implemented \cite{ar:zvs}. For the sake of completeness, this paper discusses how the results apply 
to both gate sets. Circuit depth, as defined above, upper bounds a possibly lower circuit runtime
in cases when next logic level can be executed based upon the availability of qubits and before
execution of the gates from the previous level has completed. Practically, this means 
that some of our upper bounds may not be tight (which is advantageous in the sense that the  
implementation we construct may in fact have a smaller runtime than predicted by the 
formulas evaluating depth). 
 
Quantum algorithms and their circuits are usually formulated without considering 
the physical limitations imposed by different architectures. We believe that circuit and algorithm 
designs need to be modified to account for possible architectures.
In particular, in realistic architectures, it is not possible to establish direct 
interactions between every pair of qubits \cite{ar:nmrd, ar:s-k, ar:vsb}.
A study of quantum computing architectures for the existing 
and emerging quantum technologies shows that the fastest possible direct interactions form 
a bounded degree graph ({\em e.g.}, liquid NMR quantum information processing), 
and 1D or 2D (sub)lattices \cite{ar:mo}. A mixed architecture, where values of 
stationary qubits may be teleported with the help of flying qubits to where they are desired was 
studied in \cite{ar:coi}.  In this work, the role of stationary qubits is played by 
the spins of phosphorus atoms embedded in silicon, known as the Kane proposal 
\cite{ar:sdk}, and the flying qubits are photons, with the information being teleported via 
EPR states.  Other proposals for state transfer between either stationary or 
both flying and stationary qubits, and discussions of mixed architectures, 
can be found, for example, in \cite{ar:sko, cond-mat/0603119, ar:sl} and the references therein. 
However, an architecture that allows interconversion between stationary and flying qubits 
cannot in general be realized in any technology. In addition, 
it was shown that teleportation of a {\em single} value 
(simultaneous teleportation of many qubits may be less efficient) in the Kane architecture
is only efficient if compared to {\em more} than 2-4 levels of SWAPs \cite{ar:coi}. A similar effect
is likely to take place in other mixed architecture proposals. The latter is important 
for this work since we are only using depth-1 swapping of multiple qubits via SWAP gates.

Generally speaking, due to the spatial constraints it seems unrealistic to believe that 
a direct scalable implementation of the unrestricted (where every two qubits are 
neighbors) architecture, or, more generally, unbounded neighbor architecture, will ever be found. 
Furthermore, in classical computation the number of neighbors is limited, and there is no 
obvious reason to believe that the quantum world is different. Thus, the complexity of 
the circuit designs must be refined to take it into account the limitations of possible 
quantum computing architectures. 

The linear nearest neighbor (LNN) architecture, also known as chain nearest 
neighbor, is often considered as a good (and, in fact, 
very restrictive) approximation to what a scalable 
quantum architecture may be. Mathematically, in an LNN architecture with $n$ 
qubits $q_1,\;q_2,\ldots\;,q_n$, two-qubit gates are allowed between any qubits whose
subscript values differ by one. The LNN architecture describes 1D 
lattices. It misses possible direct interactions in 2D lattices and may restrict the 
number of useful interactions in connected graphs. However, if one can show that a circuit
can be efficiently reorganized to be executed in the LNN architecture, such a circuit could 
be run efficiently in many other architectures. 

The Quantum Fourier Transformation (QFT) is an analogue of the classical discrete Fourier transformation, however, 
in the quantum case the transformation is applied to the amplitudes. The QFT serves as a basis for 
a number of efficient quantum algorithms. Most notably, it is at the heart of integer factorization and the
discrete logarithm polynomial time quantum algorithms \cite{ar:s}. Therefore, efficient implementation
of the QFT is important. This is why this topic has been studied extensively \cite{co:cw,tr:c,tr:mn}. 
Researchers presented linear and logarithmic depth circuits using a number of auxiliary qubits. 
Known circuits for the QFT have a regular structure \cite{bk:nc,tr:mn}.
However, they require direct interaction between every two qubits, which makes such circuits especially inconvenient 
for quantum architectures where only a finite number of neighbors is allowed. In an architecture 
with a finite number of neighbors, such as LNN, state transfer down the chain may require up to $(n-1)$ 
SWAP gates. We refer to the this observation as the {\em locality constraint} in the 
discussions involving lower bound arguments. A linear depth QFT 
circuit implemented in the LNN architecture has been reported in 
\cite{ar:fdh}. We reconstruct this circuit with our generalized technique and we also study lower bounds.

Stabilizer circuits (also known as unitary stabilizer circuits or Clifford group circuits) 
were introduced and studied for their use in the encoding, decoding and error detection 
stages of quantum error-correction codes \cite{ar:bd-vs,ar:g}. They can be defined 
as arbitrary quantum circuits composed with single-qubit Hadamard and Phase gates and 
two-qubit controlled-NOT gates. It turns out that
stabilizer circuits can be efficiently simulated \cite{ar:ag} as an 11-stage sequence 
of Hadamard (H), Phase (P) and linear reversible circuits (C) as H-C-P-C-P-C-H-P-C-P-C. Each P and 
H stage is a depth-1 computation composed with single-qubit gates. The depth of
stabilizer circuits is, thus, defined by the depth of a circuit realizing some linear 
reversible function. Efficient circuits for linear functions are, therefore, of great
importance. In this paper we show that every stage C can be parallelized to linear depth
in the LNN architecture. Thus, the entire stabilizer circuit requires at most linear time 
to be executed.

A very recent study shows that a size $s$ stabilizer circuit in an {\em unrestricted} 
architecture can be parallelized to a depth $O(\log n)$ circuit, but requires $O(s^3+n)$ 
auxiliary qubits \cite{arXiv:0704.1736}, Proposition 8.9. Combining the results of 
\cite{ar:ag,arXiv:0704.1736,ws:pmh} this gives a depth $O(\log n)$ circuit 
in unrestricted architectures using $O(\frac{n^6}{\log^3 n})$ auxiliary qubits to realize 
any stabilizer circuit. 
Since a depth $d$ circuit built on $q$ qubits in unrestricted architectures may become 
as large as depth $O(qd)$ in the LNN architecture (every depth-2 computation can be adversary
made to define the complete interaction pattern of the LNN architecture, and two depth-2 non-commuting 
stages can be defined such as to require a linear depth qubit permutation between them), the benefit
of logarithmic depth quickly disappears. However, a large amount of auxiliary qubits remains.  
Out approach thus appears more practical. 

The remainder of the paper is organized as follows. We start by introducing a concept 
of skeleton circuits and studying their properties. In Subsections \ref{ss:qft} and \ref{ss:lin} 
the lessons learned are applied to show that QFT and linear reversible/stabilizer circuits
can be parallelized to linear depth in the LNN architecture. Section \ref{sec:lb} reports 
lower bounds for a class of skeleton circuits which appears to be very important.  
Concluding remarks can be found in Section \ref{sec:c}.

\section{Skeleton circuits}\label{sec:sc}

Any quantum circuit composed with single-qubit and two-qubit gates can be thought 
of as a circuit composed of generic two-qubit operations each of which consists 
of a two-qubit gate of the initial circuit with the surrounding 
gates absorbed into it (the trivial case when only single-qubit gates are applied 
to a specific qubit throughout an entire computation is ignored as not interesting).
We call this a {\em skeleton circuit}. Obviously, the complexity of 
a skeleton circuit defines the complexity of the initial circuit (assuming 
that any two-qubit gate has a finite cost) and vice versa. 
We next study skeleton circuits of a certain type and apply the lessons learned 
to construct circuits for QFT and linear reversible/stabilizer circuits of linear
depth in the LNN architecture. 

\begin{figure*}
\begin{center}
\includegraphics[height=55mm]{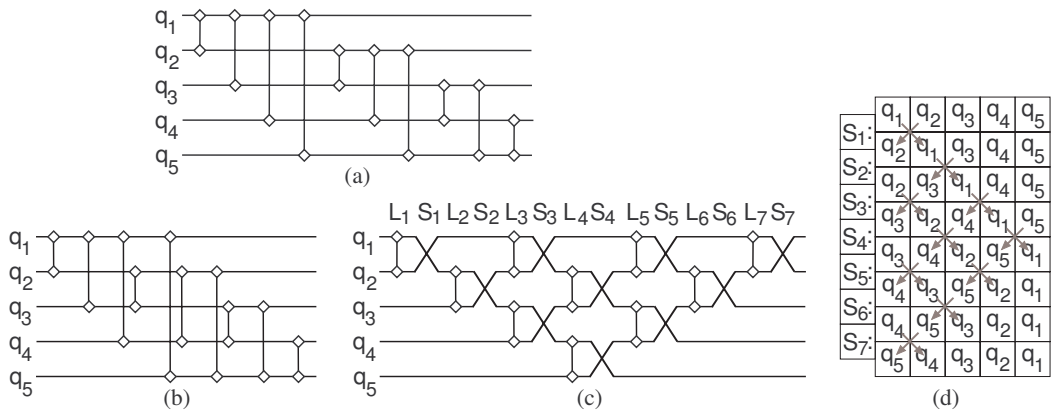}
\caption{Reorganizing an $n$-qubit skeleton circuit, illustrated for $n=5$. 
(a) Original circuit with at most $\frac{n(n-1)}{2}$ gates. Each of the gates in 
this skeleton circuit may or may not be present. (b) Linear $(2n-3)$ 
depth circuit possible to run in the ``sea-of-qubits'' architecture. (c) Version 
of (b) ready for execution in the LNN architecture.  
(d) This table illustrates how swapping stages $S_*$ are
constructed and inserted between the computational stages $L_*$.}
\label{qft5}
\end{center}
\end{figure*}

The basic skeleton circuit we consider is illustrated in Fig. \ref{qft5}(a). 
Mathematically, the skeleton circuit $SC$ is defined as 
\begin{eqnarray}\label{sc}
SC:=G_1^{i_1}(q_1,q_2) G_2^{i_2}(q_1,q_3) \ldots G_{n-1}^{i_{n-1}}(q_1,q_n) \;\;\;\;\;\;\;\;\;\; \nonumber\\
G_n^{i_{n}}(q_2,q_3)\ldots G_{n(n-1)/2}^{i_{n(n-1)/2}}(q_{n-1},q_n),
\end{eqnarray}
where $G_*$ ($*$ is reserved to represent any possible existing value of subscript) is a two-qubit gate 
that operates on the qubits indicated in brackets, $i_*$ take Boolean values, 
and for a gate $G$, $G^1$ is the gate $G$ itself, whereas $G^0=Id$ (identity, {\em i.e.},
this gate is not applied).
In other words, $i_*$ are used to indicate whether a gate is present or not. 

Since all quantum gates that operate on non-intersecting sets of qubits 
commute, the $SC$ circuit can be executed in parallel in $(2n-3)$ computational
stages $L_1,L_2,\ldots,L_{2n-3}$ defined as follows: $L_1:=G^{i_1}_1$, $L_2:=G^{i_2}_2$,
$L_3:=G^{i_3}_3 G^{i_n}_n$, $L_4:=G^{i_4}_4G^{i_{n+1}}_{n+1}$, 
$L_5:=G^{i_5}_5G^{i_{n+2}}_{n+2}G^{i_{2n-2}}_{2n-2},\ldots,$ 
$L_{2n-3}:=G^{i_{n(n-1)/2}}_{n(n-1)/2}$.
This is illustrated in Fig. \ref{qft5}(b) in the case $n=5$.

Next, the circuit can be adapted to the LNN architecture through inserting 
SWAP gates SWAP$(q_s,q_t)$ after each gate $G^{i_k}_k(q_s,q_t)$. This is 
illustrated in Fig. \ref{qft5}(c) and (d) in the case $n=5$. In the gate library 
containing all possible 2-qubit unitaries, the upper bound for depth is 
$(2n-3)$. We next use this result to achieve linear depth circuits for QFT 
and stabilizer circuits. These are fairly tight upper bounds. 
With the best known asymptotic result requiring $\Theta(n^2)$ 
gates for the QFT, it can be shown that QFT cannot be 
computed in less than linear depth even in an unrestricted architecture.  
A counting argument applied to linear circuits \cite{ws:pmh} shows that there 
exists a stabilizer circuit that requires at least $\Theta(\frac{n^2}{\log{n}})$ 
gates, meaning that it is impossible to find a circuit for it with depth less 
than $\Theta(\frac{n}{\log{n}})$ even if the architecture is unrestricted. 
Lower bounds in restricted architectures (all of which turn out to be linear, 
and thus having the same asymptotic as the upper bound that follows from our 
construction) are studied in Section \ref{sec:lb}.

Let us note that the skeleton circuit that we consider can be parallelized to 
linear depth in the LNN architecture for any initial permutation of the input and 
return the output in any desired order. For that, at most a linear depth swapping stage 
before and after the circuit is required, which does not change 
the overall linearity of the depth. The circuit illustrated in Fig. \ref{qft5}(c) 
not only allows execution in the LNN architecture, it also does not 
change the LNN connectivity pattern ($q_1-q_2-\ldots-q_n$), and thus such 
circuits can be applied one after the other with no swapping in between.
This observation will be used in Subsection \ref{ss:lin}. If the circuit 
in Fig. \ref{qft5}(c) is the last computational stage before the measurement 
is done, the last SWAP need not be applied.

\subsection{QFT in the LNN architecture}\label{ss:qft}

\begin{figure}
\begin{center}
\includegraphics[height=54mm]{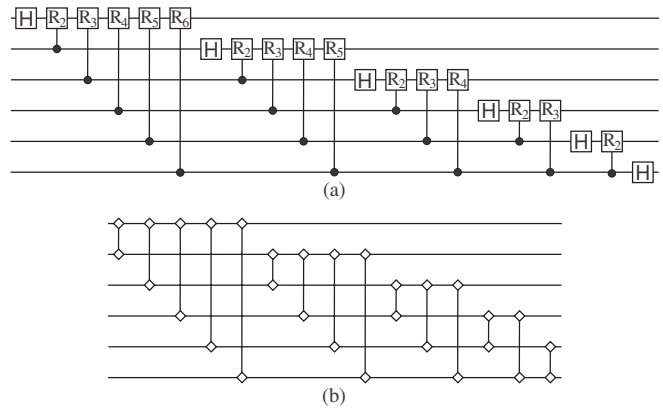}
\caption{(a) Circuit for $n$-qubit QFT \cite{bk:nc}, page 219, illustrated 
for $n=6$. The two-qubit gates are controlled-$Z$ rotations 
with parameter $1/2^{k}$, where $k$ is the subscript in the gate notation. The single-qubit 
gates are Hadamard gates. (b) Skeleton circuit of the QFT circuit in (a) composed 
of generic two-qubit gates.
}
\label{qft}
\end{center}
\end{figure}

A circuit that realizes the QFT and requires no ancilla qubits is illustrated
in Fig. \ref{qft}(a). Its skeleton circuit (Fig. \ref{qft}(b)) is obviously of 
the type considered in the previous section with all $i_*=1$. Therefore, the 
QFT can be parallelized to linear depth. This is, however, a known result, as \cite{ar:fdh} reports 
a construction that is equivalent to ours. It can also be observed that 
the approximate QFT circuit, where controlled rotations of the QFT circuit with 
small parameters are ignored, may be executed in linear depth 
in the LNN architecture. Lower bounds are discussed in Section \ref{sec:lb}, and 
they apply directly to the QFT circuit.

\subsection{Stabilizer/linear circuits}\label{ss:lin}

Synthesis of efficient linear circuits has been studied in \cite{ws:pmh}. The authors 
report a synthesis algorithm capable of producing a circuit with 
$O(\frac{n^2}{\log{n}})$ CNOT gates. It was also proven that their synthesis is 
asymptotically optimal in that there exists a linear function that requires 
$\Theta(\frac{n^2}{\log{n}})$ CNOT gates. In this paper, the goal is different.
We target minimization of the depth as opposed to the number of gates used. 
The depth of our circuit is linear in the number of qubits $n$, and it is upper bounded 
by $18n+O(1)$ CNOTs (assuming every SWAP is substituted with a suitable 3-CNOT 
implementation) or $6n+O(1)$ generic two-qubit gates. We also prove asymptotic 
optimality, which in our case is straightforward.

Every reversible linear function of $n$ variables $\vec{q}=(q_1,q_2,\ldots,q_n)^t$ can be written 
as matrix multiplication $A\vec{q}$, where $A$ is an $n \times n$ Boolean non-singular matrix. Synthesizing 
such a function is equivalent to composing a sequence of gate operations that transforms matrix
$A$ into its reduced echelon form. Due to reversibility, the reduced echelon form of $A$ is
the identity matrix. A standard technique for transforming a matrix $A$ to the identity is
to apply the Gauss-Jordan elimination algorithm. In the following, we illustrate the application 
of the Gauss-Jordan elimination algorithm and then modify its circuit to allow it be executed with 
a linear number of computational stages. Parameters $i_*$ and $p_*$ take Boolean values and they are used 
to indicate whether the gate has been applied (1) or not (0). Parameters $p_*$ are reserved for 
the gates applied to update values of the diagonal elements of the matrix $A$ during Gauss-Jordan 
elimination.

\begin{itemize}
\item Step 1. Make sure that the pivot element $a_{1,1} \neq 0$. If $a_{1,1} \neq 0$
assign $p_1:=0$. Otherwise choose $a_{j,1} \neq 0$, apply gate CNOT$(q_j,q_1)$ and 
make assignment $p_1:=1$.  
\item Steps $s=2..n$. Transform each $a_{s,1}$ to 0 through application (if needed) 
of the gate CNOT$(q_1,q_s)$. If at step $s$ a gate was applied set $i_s:=1$, otherwise,
$i_s:=0$. 
\item Step $n+1$. Make sure that the pivot element $a_{2,2} \neq 0$. If $a_{2,2} \neq 0$
do nothing ($p_2:=0$), otherwise choose $a_{j,2} \neq 0$, apply gate CNOT$(q_j,q_2)$ 
and set $p_2:=1$.
\item Steps $s=(n+2)..(2n-1)$. Transform each $a_{s,2}$ to 0 through application (if needed) 
of the gate CNOT$(q_2,q_{s-n+1})$. If at step $s$ a gate was applied set $i_s:=1$, otherwise,
$i_s:=0$. \vspace{2mm}
\newline  $\ldots$
\item Step $\frac{n(n+1)}{2}-2$. Make sure that the pivot element $a_{n-1,n-1} \neq 0$. If $a_{n-1,n-1} \neq 0$
do nothing ($p_{n-1}:=0$), otherwise apply gate CNOT$(q_n,q_{n-1})$ and make assignment 
$p_{n-1}:=1$. After this step, all parameters $p_*$ must be set.
\item Step $\frac{n(n+1)}{2}-1$. Transform each $a_{n,n-1}$ to 0 through application (if needed) 
of the gate CNOT$(q_{n-1},q_n)$. If the gate was applied set $i_{\frac{n(n+1)}{2}-1}:=1$, otherwise,
$i_{\frac{n(n+1)}{2}-1}:=0$. At this point, the set of applied transformations reduced
matrix $A$ to the upper triangular form with ones on diagonal. The remainder of the algorithm 
eliminates non-zero elements above the diagonal. 
\item Steps $s=\frac{n(n+1)}{2}..(n^2-1)$. If $a_{k,l} \neq 0$, apply CNOT$(q_l,q_k)$ for $k=l..1$ 
inside for $l=n..2$ and set $i_s$ to one iff a gate has been applied. 
\end{itemize}

\begin{figure}
\begin{center}
\includegraphics[height=53mm]{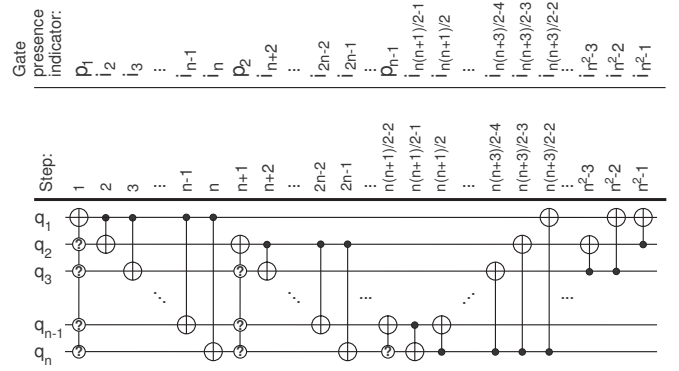}
\caption{Application of Gauss-Jordan elimination algorithm to the synthesis of a reversible network. 
Gates with controls $\bigcirc\!\!\!\!\!\!\!{\tiny\bf ?}$\;\; indicate a single CNOT each with the control 
at (exactly) one of positions marked $\bigcirc\!\!\!\!\!\!\!{\tiny\bf ?}$\;\;.} 
\label{gj}
\end{center}
\end{figure}

\begin{figure}
\begin{center}
\includegraphics[height=42mm]{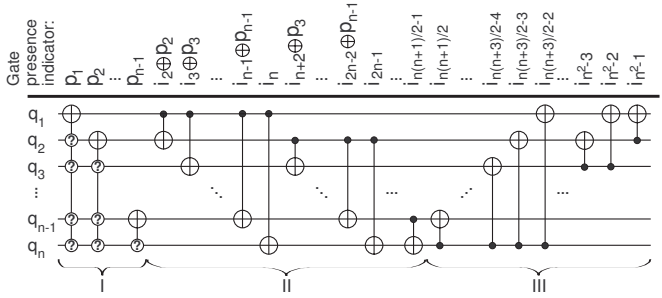}
\caption{Gauss-Jordan elimination algorithm network with rearranged gates.}
\label{gj2}
\end{center}
\end{figure}

\begin{figure}
\begin{center}
\includegraphics[height=9mm]{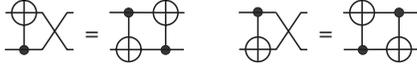}
\caption{2-CNOT circuit equivalent to a SWAP-CNOT pair.}
\label{swapcnot}
\end{center}
\end{figure}

We next use the gate commutation rule (two CNOT gates commute iff target of one gate is not equal to
the control of the other) and circuit identity CNOT$(a,c)$CNOT$(c,b)=$ CNOT$(c,b)$CNOT$(a,b)$CNOT$(a,c)$ 
to move all $(n-1)$ gates CNOT$(a,c)$ with parameter $p_*$ to the front of the network. Note, that every time 
commutation rule is used, the gates just change their position and every time the circuit identity 
is applied we introduce a new gate CNOT$(a,b)$. However, such a gate can always be commuted to the closest
on the left CNOT$(a,b)$, and this is accounted for by the updates to the $i_*$ gate presence indicator.
The circuit gets transformed to the one illustrated in Fig. \ref{gj2}. Parameters $i_*$ are changed
through XORing each $i_j$, $j < \frac{n(n+1)}{2}$ with $p_k$, for $k < n$ such that $q_k$ is the target 
of the gate used at step $j$. The constructed circuit consists of three parts marked I-III in 
Fig. \ref{gj2}. The skeleton of each of these parts is described by equation (\ref{sc}), which is 
obvious for parts II and III and requires a short explanation for part I. Divide the skeleton circuit (Fig. \ref{qft5}a)
into $(n-1)$ parts with the first containing first $(n-1)$ gates, the second containing next $(n-2)$ gates, 
and so on, the last, $(n-1)^{\textrm{st}}$ part containing one last gate. Then, gate $G_i$ for $i=1..n-1$ from 
part I of the circuit in Fig. \ref{gj2} can be matched (via ``skeletonization'') to some gate 
in the $i^{\textrm{th}}$ part of the skeleton circuit $SC$. Thus, every linear reversible function can 
be computed as a maximal depth $3(2n-3)=6n+O(1)$ circuit. Furthermore, since each SWAP-CNOT
pair can be rewritten as two CNOTs (Fig. \ref{swapcnot}) and SWAP requires no more than 3 CNOT gates,
the overall depth in terms of CNOTs can be upper bounded by the expression $18n+O(1)$. We note that in 
some quantum information processing proposals pair CNOT-SWAP can be executed more efficiently than 
a single CNOT or a single SWAP, such as in \cite{ar:fhh}, Fig. 1.  Due to the locality 
constraint our upper bound has the same asymptotic as a lower bound, and thus our circuits are 
asymptotically optimal. Using H-C-P-C-P-C-H-P-C-P-C decomposition 
for stabilizer circuits \cite{ar:ag} these upper bounds directly translate to at 
most depth $30n+O(1)$ circuit composed with generic two-qubit gates, or at most depth 
$90n+O(1)$ circuit in the library with single-qubit and CNOT gates.  

\subsubsection{Encoding and error syndrome circuits for CSS codes}

\begin{figure}[ht]
\begin{center}
\includegraphics[height=40mm]{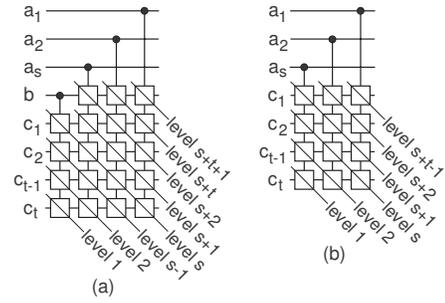}
\caption{General structure of the (a) encoding and (b) syndrome detection circuits for CSS quantum error 
correcting codes.}
\label{ecc}
\end{center}
\end{figure}

Encoding and error syndrome circuits for CSS codes are of a great practical importance due to the 
clever error correcting properties of the CSS codes. Such circuits include those 
illustrated in Fig. \ref{ecc}(a) (encoding; \cite{co:mtc}) and Fig. \ref{ecc}(b) (error 
syndrome; \cite{bk:nc}), where single-qubit Hadamard gates are not illustrated
since their contribution to the total depth is only a constant, and the controlled gates, each of which 
may or may not be present (which is defined by the form of the parity check matrices of the corresponding 
classical codes), are either controlled-NOT or controlled-Z. Our circuit parallelization technique 
described in the previous subsection applies directly to such circuits
since each of them has skeleton as described by the Eq. (\ref{sc}) with 
$n=s+t+1$ for the encoding circuit and $n=s+t$ for the error syndrome circuit. This allows us to execute 
the encoding circuit in $(2s+2t-1)$ stages and the error syndrome circuit in $(2s+2t-3)$ 
(in both cases, $s,t \geq 1$) stages composed of generic
two-qubit gates. However, a better approach is possible. The following construction is, essentially,
a part of the algorithm used to execute $SC$.

Consider encoding circuit (Fig. \ref{ecc}(a)). Prepare the qubits in the following LNN connectivity pattern 
$a_1-a_2-\ldots-a_s-b-c_t-c_{t-1}-\ldots-c_1$. At each level $i$ apply gates whose targets intersect with the 
sloping lines marked ``level $i$'' shown in Fig. \ref{ecc}(a). Each such level is followed by 
the level of SWAPs applied to the same qubits as the gates from the previous level to allow for the next
set of gates to get executed in the LNN architecture.
For example, for $s=3$ and $t=4$ level 3 will be composed of the gates $G(b,c_2)$, $G(a_3,c_3)$, 
and $G(a_2,c_4)$, followed by the swaps SWAP$(b,c_2)$, SWAP$(a_3,c_3)$, and SWAP$(a_2,c_4)$. Thus, 
the total depth of the encoding circuit executable in the LNN architecture 
will be equal to $(s+t+1)$ if it is allowed to be composed of generic two-qubit gates. 
This is almost half of what was expected if this circuit were matched to the $SC$ first. 
This translates to a depth $2(s+t+1)$ circuit with controlled-NOT, controlled-Z and SWAP gates. 
Similarly, the depth of the error syndrome circuit composed with generic gates and executable in 
the LNN architecture is $(s+t-1)$.

Application of the technique described in this subsection to executing the error syndrome circuit 
for Steane's code (\cite{bk:nc}, Fig. 10.16) in the LNN architecture shows that this can be done 
in 12 stages composed of generic two-qubit gates or 26$(=2*12+2)$ stages composed of Hadamard, 
controlled-NOT, controlled-Z, and SWAP gates. We can show that the encoding circuit of 
\cite{co:mtc}, Fig. 8b, can be executed in 23 stages composed of generic two-qubit gates or, alternatively, 
68 ($=3*23+1-2$: pairs CNOT-SWAP must be combined, we need an extra level for Hadamard gates, 
but do not need to apply last SWAP)
stages composed of CNOT and Hadamard gates in the LNN architecture. Our result for the   
depth, 68, is notably better than 177 found by the automated procedure of \cite{co:mtc}.

\section{Lower bounds}\label{sec:lb}

In this section we study lower bounds on the depth of skeleton circuit $SC$
defined in equation (\ref{sc}) assuming all gates are present ({\it i.e.}, each $i_*=1$). 
We further assume that a pair of gates $G(q_i,q_j)$SWAP$(q_i,q_j)$ requires two units 
of the execution time, one for each of the gates. In practice, a direct implementation 
of pair $G(q_i,q_j)$SWAP$(q_i,q_j)$ may be more efficient \cite{ar:zvs}, but the particulars 
of such a construction depend on the specific Hamiltonian, which is unknown in the general case. 
The depth of circuit illustrated in Fig. \ref{qft5}(c) is thus $(4n-6)$. The lower 
bounds achieved below are directly applicable to the QFT circuit. 

To prove lower bounds, we need to restrict the set of possible computations.
We define two circuit type quantum computational models $A$ and $B$. We require
that for each of them in order to compute the $SC$ (equation (\ref{sc})) all $\frac{n(n-1)}{2}$ two-qubit gates 
need to be executed, and no ancilla qubits may be used. Furthermore,
\begin{itemize}
\item in model $A$ we assume that the gates required to be executed in $SC$ cannot be commuted
(other than trivially---a pair of gates operating on non-intersecting sets of qubits always commutes);
\item in model $B$ we allow possibility of the execution of gates in any order ({\it i.e.}, this 
lets us obtain bounds that allow commuting gates through the circuit, without 
worrying about which gates actually commute, and what kind of corrections are needed in case they 
do not commute).
\end{itemize}
The architectures considered in this paper are {\em LNN}, {\em 2D square lattice}, 
and {\em bounded degree graph} with the degree of each vertex no more than $k$.
We next prove a number of lower bounds, refer to Table \ref{tab:lb}. 
\begin{table}[ht]
\caption{\label{tab:lb} Lower bounds on the depth of the $SC$ in models $A$ and $B$ in the {\em LNN}, 
{\em 2D square lattice}, and {\em bounded degree graph} architectures.}
\begin{ruledtabular}
\begin{tabular}{cccc} \hline
    & {\bf LNN} & {\bf 2D square lattice} & {\bf bounded degree graph} \\ \hline
model $A$ & $\frac{10n}{3} + O(1)$ & $3n + O(1)$ & $(2+\frac{2}{k})n + O(1)$ \\
model $B$ & $\frac{3n}{2} + O(1)$ & $\frac{5n}{4} + O(1)$ & $(1+\frac{1}{k})n + O(1)$ \\ \hline
\end{tabular}
\end{ruledtabular}
\end{table}

\noindent {\bf $\frac{10n}{3} + O(1)$ bound in LNN, model $A$.}
First, denote each depth-1 computational stage (logic level) by $L$ and each depth-1 swapping stage by $S$.
Every three stages of the $SC$ have a single fixed qubit that interacts 
with three other qubits. This is either $q_1$, $q_2$, or $q_n$. Thus, every three logic levels 
have to be separated by a round of SWAPs, each having depth at least 1, {\em i.e.} 
each sequence $LLL$ must be replaced by $LSLL$ or $LLSL$ to be able to run the circuit in the LNN 
architecture. We call this $3L\rightarrow 1S$ requirement. With the $3L\rightarrow 1S$ requirement, the total 
depth must be at least $2n-3+\lceil\frac{1}{2}(2n-5)\rceil = 3n + O(1)$ logic levels. Therefore, using 
just the $3L\rightarrow 1S$ requirement proves that our circuit is at most factor $\frac{4}{3}$ 
off the optimum. We now improve this bound to $\frac{10n}{3} + O(1)$ by showing that 
every 4 computational stages must be separated by at least depth-2 swapping stage ($4L\rightarrow 2S$ requirement). 
$4L\rightarrow 2S$ is slightly more restrictive than $3L\rightarrow 1S$. The difference between the two is
that in one $LLSLL$ is allowed, but not in the other. We next prove that depth-1 level 
does not suffice in separating some two computational stages from the following two by 
exploring the properties of $SC$ and the LNN architecture.

Assume all 4 computational stages $L_i,\;L_{i+1},\;L_{i+2}$, and $L_{i+3}$ are solely in the first half of $SC$. 
The second half is symmetric to the first half and thus a similar proof holds for it. We do not 
prove the boundary case (where one part of the 4-stage computation is in the first half of the $SC$
and the other part is in the second half) because its contribution to the final figure is only a 
constant. Next, assume $i$ is odd. The proof for even values $i$ is analogous. 
Name the qubits $q_1,q_2,\ldots,q_n$ top to bottom. The computational stages $L_i$ and $L_{i+1}$ 
use interactions $q_{i+2}-q_1$, $q_1-q_{i+1}$, $q_{i+1}-q_2,\ldots$ $,q_{\frac{i+1}{2}}-q_{\frac{i+3}{2}}$,
which in the LNN architecture can only be aligned as follows: 
$q_{i+2}-q_1-q_{i+1}-q_2-\ldots-q_{\frac{i+1}{2}}-q_{\frac{i+3}{2}}$.
The computational stages $L_{i+2}$ and $L_{i+3}$ use interactions 
$q_{i+4}-q_1$, $q_1-q_{i+3}$, $q_{i+3}-q_2,\ldots$ $,q_{\frac{i+3}{2}}-q_{\frac{i+5}{2}}$. 
In particular, stages $L_{i+2}$ and $L_{i+3}$ require interaction $q_{\frac{i+3}{2}}-q_{\frac{i+7}{2}}$, 
and qubit $q_{\frac{i+7}{2}}$ is used both in $L_{i+2}$ and $L_{i+3}$. 
However, we know that after completion of stages $L_i$ and $L_{i+1}$, the architecture allows 
interactions in the following order 
$q_{\frac{i+3}{2}}-q_{\frac{i+1}{2}}-q_{\frac{i+5}{2}}-q_{\frac{i-1}{2}}-q_{\frac{i+7}{2}}$.
The LNN architecture distance between $q_{\frac{i+3}{2}}$ and $q_{\frac{i+7}{2}}$ is 4. 
A depth-1 swapping reduces the architectural distance between these qubits by at most 2, which is not enough 
for the desired interaction to be allowed. Thus, the depth of swapping must be at least 2. This 
concludes the proof of the $4L\rightarrow 2S$ requirement.
  
We finalize the proof of $\frac{10n}{3} + O(1)$ lower bound by observing that  
for a circuit with $2n + O(1)$ stages $L$ we need to have at least $\frac{4n}{3} + O(1)$ stages
$S$ to satisfy $4L\rightarrow 2S$ requirement. Thus, the total number of stages required 
to execute $SC$ in LNN is $\frac{10n}{3} + O(1)$.
This implies that the circuit we constructed explicitly 
(Fig. \ref{qft5}(c)) must be within factor of $\frac{6}{5}$ from optimum. \\

\noindent {\bf $3n + O(1)$ lower bound in 2D square lattice, model $A$.}
We prove that every three computational stages $L_{i-2},\;L_{i-1}$, and $L_{i}$, where $i=2k+1$ and 
$k=1..\lceil \frac{n-2}{2} \rceil$ (this means that all computational stages are in the 
first part of $SC$; the proof for the symmetric second part is similar) must contain at 
least one swapping stage if ran in 2D square lattice architecture. We prove this by finding
three interactions that form a loop. Vertices in such loop cannot be isomorphically mapped 
to the vertices of 2D square lattice. The interactions that form such a loop, assuming 
qubits are named $q_1, q_2,\ldots, q_n$ top to bottom, are 
$q_{\frac{i-1}{2}}-q_{\frac{i+1}{2}}$ in $L_{i-2}$, $q_{\frac{i-1}{2}}-q_{\frac{i+3}{2}}$ 
in $L_{i-1}$, and $q_{\frac{i+1}{2}}-q_{\frac{i+3}{2}}$ in $L_{i}$. This proves that 
for every possible value $k$ it is required to have at least one swapping stage, which 
results in the construction of $3n + O(1)$ lower bound. 

The lower bound that we just proved may be interesting to those experimentalists 
working on implementing 2D architectures for quantum information processing. The lower bound 
shows that, with certain restrictions, the QFT in 2D square lattices cannot in principle 
be parallelized any more efficiently than to a depth at least $\frac{3}{4}$ of the depth of QFT 
circuit executable in the LNN architecture. \\

\noindent {\bf $\frac{3n}{2} + O(1)$ lower bound in NCT, model $B$.}
Recall that the number of gates in $SC$ is $\frac{n(n-1)}{2}$ and they all require different qubit-to-qubit
interactions to be available. Next, note that in the LNN architecture application of a single 
SWAP may make at most two new interactions become available for a gate 
to be applied on. Thus, the total number of SWAPs that one must execute in a 
circuit to go through all $\frac{n(n-1)}{2}$ possible interactions is at least 
$\lceil\frac{\frac{n(n-1)}{2}-(n-1)}{2}\rceil=\lceil\frac{(n-1)(n-2)}{4}\rceil$. This means that
the total number of gates to be executed in the LNN architecture to compute $SC$ must be at least  
$\frac{n(n-1)}{2}+ \lceil\frac{(n-1)(n-2)}{4}\rceil = \lceil\frac{(3n-2)(n-1)}{4}\rceil$.
At most $\lfloor\frac{n}{2}\rfloor$ gates can be executed in parallel. Thus, the depth of the circuit is at least
the minimum total number of gates to be executed divided by the maximum number of gates that can be 
executed simultaneously, {\em i.e.} $\frac{3n}{2} + O(1)$.  

This lower bound is constructed based on the assumption that all gates in $SC$ need 
to be executed, and does not take into account that the order they are executed
in is important. Thus, the restriction on the form of the computation is significantly weaker than 
that for model $A$, and the proven lower bound is looser. \\

Generalizing the above techniques, it can be shown that in an architecture where each qubit has 
a finite number of neighbors bounded by number $k$: 
\begin{itemize}
\item the lower bound for executing $SC$ is $(2+\frac{2}{k})n + O(1)$ in model $A$; 
\item the lower bound for executing $SC$ is $(1+\frac{1}{k})n + O(1)$ in model $B$.
\end{itemize}

The $\frac{5n}{4} + O(1)$ lower bound announced in Table \ref{tab:lb} follows from the second of these 
two statements. Given the linearity of proven lower and upper bounds, we have just shown the asymptotic 
optimality of the depth of our skeleton circuit in the restricted architectures considered
in this paper.

\section{Conclusion}\label{sec:c}

In this paper we studied the complexity of the execution of the quantum Fourier transformation and 
stabilizer circuits in restricted architectures.

We reconstructed the depth $4n+O(1)$ circuit (composed with SWAP and controlled-$Z$ gates)
for QFT initially reported in \cite{ar:fdh} which is implemented 
in the LNN architecture. With the application of our generalized technique we showed
how the approximate QFT circuit can be executed in linear depth in the LNN architecture. 
We proved a number of lower bounds for the depth of QFT circuit, 
which are all a constant factor away (ranging from $\frac{1}{4}$ to $\frac{5}{6}$, and depending 
on the computational model and assumptions made) from the above upper bound. Some of our lower 
bounds can be used by experimentalists working on implementing advanced architectures as a guide 
to how complex architectures may need to be for particular types of computations. For instance, 
we proved that, with certain restrictions, the QFT circuit in 2D square lattices cannot in principle 
be parallelized more than to the depth equal to $\frac{3}{4}$ of the depth of QFT circuit 
executable in the LNN architecture. 

More importantly, we presented a constructive algorithm for synthesizing linear depth stabilizer circuits 
in the LNN architecture. In particular, we showed that any stabilizer circuit can be executed 
in at most $30n+O(1)$ stages each composed with generic two-qubit gates, which in the library with CNOT and 
single-qubit gates translates to at most depth $90n+O(1)$ circuit. This upper bound is
asymptotically optimal. We considered specific stabilizer circuits and showed how 
these circuits can be executed faster than reported by previous researchers \cite{co:mtc}.

\section*{Acknowledgments}

I would like to thank Prof. Michele Mosca from the University of Waterloo and for his help in preparation 
of this manuscript and useful discussions. I wish to thank Jacob D. Biamonte from the University of Oxford and 
Donny Cheung from the University of Waterloo for their help in preparation and proofreading this manuscript.  
This work was supported by PDF grant from the National Sciences and Engineering Research Council of Canada.

\end{document}